\documentclass[epj]{svjour}
\usepackage{mathrsfs}
\usepackage{amsmath}
\usepackage{amsfonts}
\usepackage{amssymb}
\usepackage[numbers]{natbib}
\usepackage{color}
\usepackage[caption=false]{subfig}
\usepackage{graphicx}
\usepackage{hyperref}

\begin{document}

\title{An all-electrical torque differential magnetometer operating under ambient conditions}

\author{Akashdeep Kamra\inst{1,2} \and Stefan von Hoesslin\inst{1} \and Niklas Roschewsky\inst{1} \and Johannes Lotze\inst{1} \and Michael Schreier\inst{1} \and Rudolf Gross\inst{1,3,4} \and Sebastian T. B. Goennenwein\inst{1,3} \and Hans Huebl\inst{1,3} \thanks{email: huebl@wmi.badw.de}}

\institute{Walther-Mei{\ss}ner-Institut, Bayerische Akademie der Wissenschaften, Walther-Mei{\ss}ner-Str. 8, 85748 Garching, Germany \and Kavli Institute of NanoScience, Delft University of Technology, Lorentzweg 1, 2628 CJ Delft, The Netherlands \and Nanosystems Initiative Munich (NIM), Schellingstr. 4, 80799 Munich, Germany \and Physik-Department, Technische Universit\"at M\"unchen, James-Franck-Str. 1, 85748 Garching, Germany}

\date{Received: date / Revised version: date}

\abstract{
An all-electrical torque differential magnetometry (also known as cantilever magnetometry) setup employing piezoelectric quartz tuning forks is demonstrated. The magnetometer can be operated under ambient conditions as well as low temperatures and pressures. It extends the allowed specimen mass range up to several 10 $\mu$g without any significant reduction in the sensitivity. Operation under ambient conditions and a simple all-electrical design of the magnetometer should allow for an easy integration with other experimental setups. The uniaxial magnetic anisotropy of a 25 $\mu$m diameter iron wire, measured under ambient conditions with a high signal to noise ratio, was found to be in good agreement with its literature value. Further applications of the technique are discussed.
}

\maketitle

%-------------------------------------------------------------------------------------------------------%

\section{Introduction}
Cantilever mechanical resonators have found application in a wide range of sensing and detection schemes including mass sensing~\cite{Moser,Lavrik}, atomic (magnetic) force microscopes~\cite{Giessiblrmp}, chemical sensors~\cite{Nordstrom}, torque magnetometry~\cite{Rossel1996,Willemin1998,Rossel1998} and torque differential magnetometry~\cite{Kamra} (also known as cantilever magnetometry~\cite{Stipe,Weber}). Most of these measurement schemes are based on detection of changes in either the oscillation amplitude at a fixed drive or the resonance frequency of the mechanical resonator. This change in amplitude or resonance frequency can in turn be attributed to a change in the effective mass ($m_{\textrm{eff}}$) or the effective spring constant ($k_{\textrm{eff}}$) of the mechanical resonator induced by the physical parameter to be detected (henceforth called {\it perturbation}). Thus, to achieve a high sensitivity, small $m_{\textrm{eff}}$ and $k_{\textrm{eff}}$  are required to obtain a large frequency shift for a given perturbation. The trade-offs include 
small specimen size and exclusive operation at low temperatures and pressures. Further, the resonant response of these resonators is typically recorded using a mechanical piezoelectric drive and optical detection of the oscillation amplitude~\cite{Stipe}. This optical detection makes the setup relatively expensive and sensitive to external disturbances. 

Complementary, purely electrical detection schemes have also been investigated including piezoresistive resonators~\cite{Tortonese,Eriksson} and piezoelectric quartz tuning forks	 (TFs)~\cite{Giessiblnanotech,Rychenrsi,Todorovic,Unterreithmeier}. The latter resonators are particularly attractive as they are commercially available and can be integrated on small footprints. In addition to high quality factors of about 10000 under ambient conditions, they offer several other desirable properties like high temperature stability, low sensitivity to external mechanical disturbances, and robustness~\cite{Bottom}. Further, the relatively large $k_{\textrm{eff}}$ of these resonators offers several advantages due to smaller oscillation amplitude~\cite{Giessiblnanotech} and larger linear operation range~\cite{Kamra} while preserving the high sensitivity, since the drawback of smaller frequency shifts due to large $k_{\textrm{eff}}$ is compensated by a high quality factor resulting in better frequency resolution.

\begin{figure}[htb]
\centering
\includegraphics[width=8.5cm]{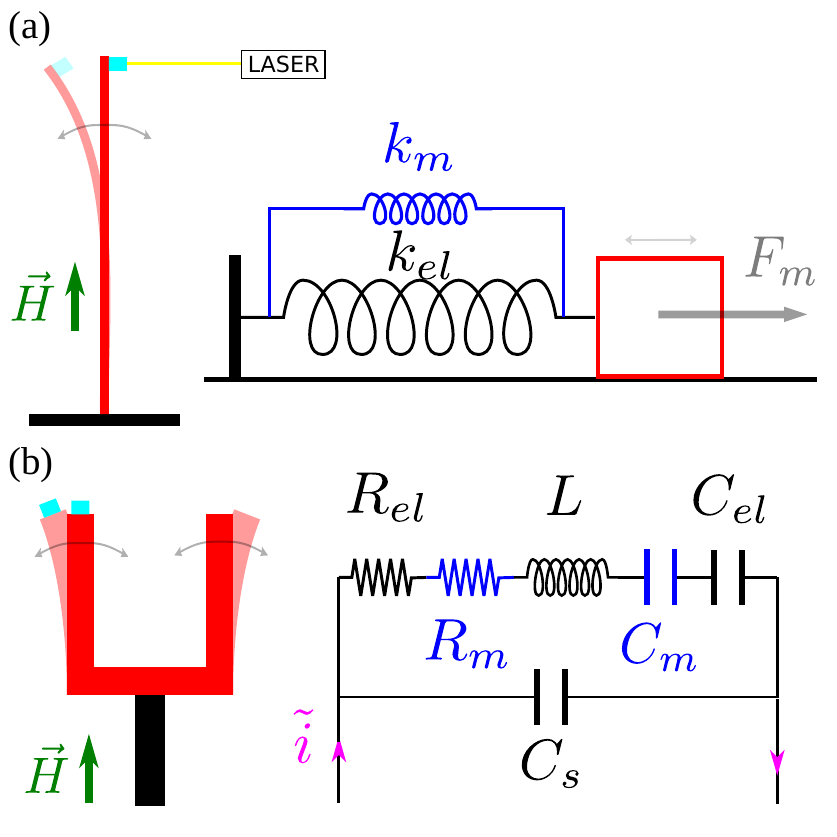}
\caption{Torque differential magnetometry~\cite{Stipe,Kamra} using a cantilever (a) and a quartz TF (b). On the rhs the mechanical and electrical equivalent circuits are shown. A magnetic specimen (light blue) is attached to the tip of the resonator (red) and the former experiences a torque under the influence of an applied magnetic field. This torque translates to a magnetic field dependent effective stiffness (capacitance) leading to magnetic field dependent resonance frequency of the cantilever (TF). The applied magnetic field dependent quantities are shown in blue. $R_m$ represents the magnetic contribution to the dissipation. The cantilever setup requires a laser interferometer for measurements while the TF enables an all-electrical measurement scheme.}
\label{ctmvstdm}
\end{figure}

We have employed quartz TFs to perform {\it quantitative} measurements via torque differential magnetometry (TDM)~\cite{Stipe,Kamra}. Our setup improves upon the existing TDM capabilities~\cite{Stipe,Weber} by allowing `large' specimens weighing up to several 10 $\mu$g, operation over a broad temperature and pressure range, higher sensitivity for detecting magnetic contribution to dissipation, and a simple all-electrical implementation. We give a brief introduction to the quartz TF as a mechanical resonator in Sec. \ref{tfresonator}. TFs have already been used for high resolution microscopy~\cite{Giessiblnanotech,Rychenrsi} and alternating gradient magnetometry~\cite{Todorovic}. However, in all these measurements, only qualitative changes in the surface topography or magnetic moment were of interest. In contrast, we demonstrate, in a proof-of-principle experiment, a quantitative analysis of the anisotropy field and magnetization of a thin iron wire with a high precision. The TDM 
method entails attaching the magnetic specimen to the tip of the TF and recording the resonance frequency as a function of an applied magnetic field (see Fig. \ref{ctmvstdm}). We detail optimal mounting procedures in Sec. \ref{setupsec} and experimental spectroscopy techniques in Sec. \ref{measurement}.

In TDM, the magnetic specimen experiences a mechanical torque, which acts as an effective force $F_m$ [see Fig. \ref{ctmvstdm} (a)], under the influence of an externally applied magnetic field. This additional restoring force translates to an additional magnetic field dependent stiffness ($k_m$). Since the resonance frequency $f_r$ of the mechanical resonator depends on the total effective stiffness constant $k_{\textrm{eff}} = k_{m} + k_{\textrm{el}}$ via $f_r = 1/2\pi \sqrt{k_{\textrm{eff}}/m_{\textrm{eff}}}$, the magnetic torque imposes a magnetic field dependent resonance frequency shift ($\delta f_r$). In a TF [Fig. \ref{ctmvstdm} (b)], the effective stiffness is inversely related to the capacitance $C = (C_m C_{\textrm{el}})$ $/ (C_m + C_{\textrm{el}})$ in the equivalent circuit, $C_m$ being magnetic field dependent, which in turn determines the resonance frequency as detailed in Sec. \ref{tfresonator}. Further, there is magnetic field dependent contribution to the dissipation represented by the 
resistance $R_m$ in the equivalent circuit. In Sec. \ref{tdmsec}, we detail how the observed frequency shift can be used to quantify the magnetic properties of the specimen~\cite{Kamra}.

%--------------------------------------------------------------------------------------------%

\section{Tuning Fork Resonator}\label{tfresonator}
Commercially available quartz TFs are designed to have a perfectly anti-symmetric resonance mode, with one prong mirroring the other, with a resonance frequency of 32768 Hz. Like any other mechanical oscillator~\cite{Morse}, piezoelectric quartz TFs can be modeled as an effective mass and spring system with friction. Due to piezoelectricity, there is a direct relation between the deflection $x$ of the effective mass and the charge accumulated across the electrodes $Q = \alpha x$, with $\alpha$ the electromechanical coupling constant. Comparing the mechanical setup with the lumped element $LCR$ equivalent electrical resonance circuit allows to identify the following relations~\cite{Rychenphd,Friedt} (see Fig. \ref{ctmvstdm}):
\begin{equation}\label{equiv}
 L = \frac{m_{\textrm{eff}}}{\alpha^2}, \quad \frac{1}{C} = \frac{k_{\textrm{eff}}}{\alpha^2}, \quad R = \frac{\gamma_{\textrm{eff}}}{\alpha^2}.
\end{equation}
Here $C^{-1} = C_{\textrm{el}}^{-1} + C_m^{-1}$, $R = R_{\textrm{el}} + R_m$, and $-\gamma_{eff} = -\gamma_{\textrm{el}} - \gamma_{m}$ is the proportionality constant between the friction force and the velocity, in the mechanical model. The negative sign emphasizes that the friction acts against the motion. In addition to the {\it motional} $LCR$ equivalent circuit, the TF has a physical shunt capacitance $C_s$, which acts in parallel to the motional branch, leading to the Butterworth-van Dyke (BvD) equivalent circuit~\cite{Rychenphd,Friedt}, as shown in the lower right panel of Fig. \ref{ctmvstdm}. The admittance $Y(\omega)$ for this circuit exhibits a resonance at $\omega_r = 2 \pi f_r = 1/\sqrt{LC}$ and an anti-resonance at $\omega_{\textrm{ar}} = 2 \pi f_{\textrm{ar}} = 1/\sqrt{L[C C_s /(C + C_s) ]}$, and is given by
\begin{eqnarray}
 Y(\omega) = \frac{\tilde{I}(\omega)}{\tilde{V}(\omega)} & = & \frac{1}{R + \frac{1}{i \omega C} + i \omega L} + i \omega C_s,
\end{eqnarray}
where $\tilde{I}$ and $\tilde{V}$ denote ac quantities. The admittance can be recast in the form of a complex Lorentzian, characteristic for any resonance, close to the resonance frequency:
\begin{eqnarray}\label{lorentzian}
 Y(f) & = &  \frac{A_0 \Delta f \left( \Delta f - 2 i (f - f_r)  \right)}{(\Delta f)^2 + 4 (f - f_r)^2} + i 2 \pi f_r C_s,
\end{eqnarray}
with $A_0 = 1/R$ and $\Delta f = R / 2 \pi L$. A high quality factor $Q = f_r/\Delta f$ is desirable as it quantifies the precision in the measurement of the resonance frequency.

\begin{figure}[htb]
\centering
\includegraphics[width=8.5cm]{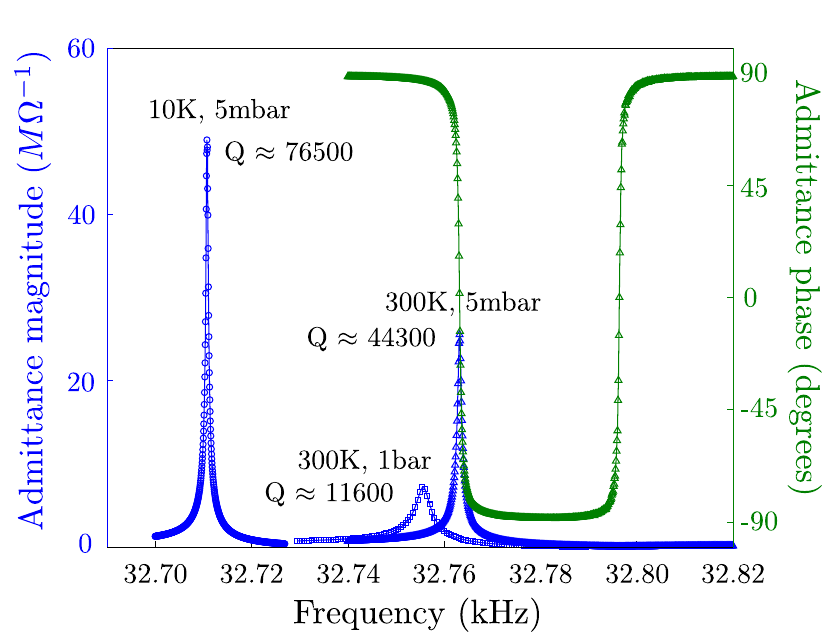}
\caption{Admittance magnitude (left axis, blue curves) and phase (right axis, green curve) for a TF removed from its casing under different environments. Only one phase curve, corresponding to $T=300$\,K and $p = 5$\,mbar, is shown to avoid crowding in the figure.}
\label{admcurves}
\end{figure}

The measured magnitude and phase of the admittance are shown in Fig. \ref{admcurves} for a TF (casing removed) at ambient conditions (squares), 300 K and 5 mbar pressure in helium environment (triangles), and 10 K and 5 mbar pressure in helium exchange gas (circles). Fitting the admittance magnitude recorded at 300 K and 5 mbar pressure with a complex Lorentzian [Eq. (\ref{lorentzian})] yields $R_{\textrm{el}} = 39 ~\textrm{k}\Omega$, $L = 8400 ~\textrm{H}$, $C_{\textrm{el}} = 2.8 ~\textrm{fF}$, and $C_s = 1.4 ~\textrm{pF}$. The quality factor was found to increase by about an order of magnitude in going from ambient conditions to 10 K and 5 mbar. Quality factors above $10^6$ have been reported at still lower temperatures and pressures~\cite{Gomez}.

%-------------------------------------------------------------------------------------------------------------------%

\section{Setup}\label{setupsec}
\begin{figure}[htb]
\centering
\includegraphics[width=8.5cm]{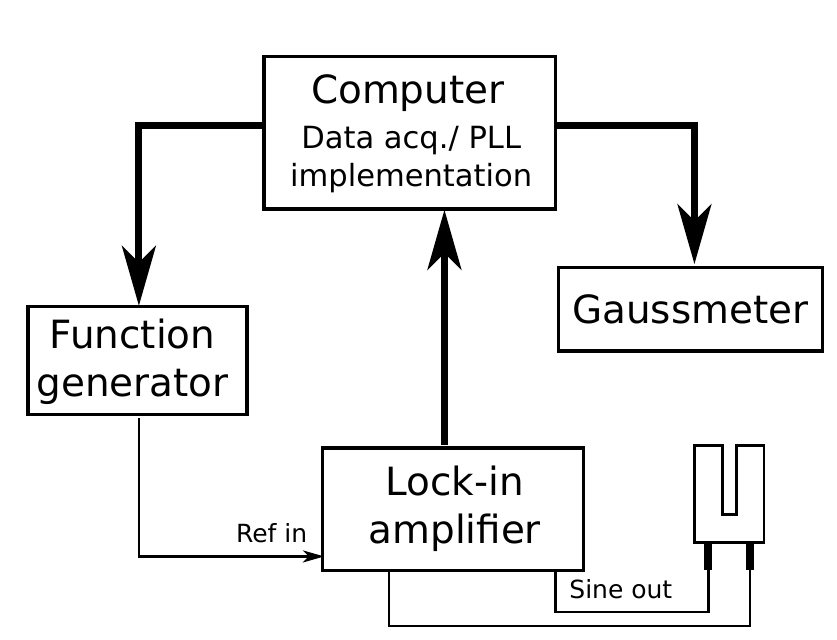}
\caption{Schematic showing the electronics used in the setup. Thick lines denote GPIB connections while the thin lines represent coaxial cables.}
\label{setup}
\end{figure}

{\it All measurements reported below were performed under ambient conditions.} In our measurements, we perform admittance spectroscopy of the device as sketched in Fig. \ref{setup}. A lock-in amplifier is used to apply a small ac voltage $\tilde{V}(\omega)=V_1 \cos \omega t$ with amplitude $V_1=4$\,mV and angular frequency $\omega$ across the TF electrodes, and simultaneously measure the current response $\tilde{I}(\omega)$. The ratio of the current response $\tilde{I}(\omega)$ and the applied ac voltage $\tilde{V}(\omega)$  gives the complex admittance $Y(\omega)$ at the given frequency. Admittance over a certain frequency range or at the resonance frequency only may be of interest as per the requirements of the measurement. 

To obtain a high frequency resolution, we employ an Agilent 33250A function generator which provides the reference signal to a Stanford Research SR830 lock-in amplifier, whose voltage output and current input ports are respectively used for applying the voltage drive and measuring the current response (see Fig. \ref{setup}). The high impedance of the TF under all conditions, except very low temperature and pressure, enables a direct current measurement using the SR830 current input port (impedance $1 ~\mathrm{k}\Omega$). A Lakeshore 455 DSP gaussmeter was used to control the magnetic field in a home build electromagnet. All data acquisition and the phase locked loop (PLL) implementation described in Sec. \ref{measurement} were achieved using Labview. Taken together, the setup allows to record the admittance of the TF as a function of the applied magnetic field.

\begin{figure}[htb]
 \centering
 \subfloat[]{\includegraphics[height = 5cm]{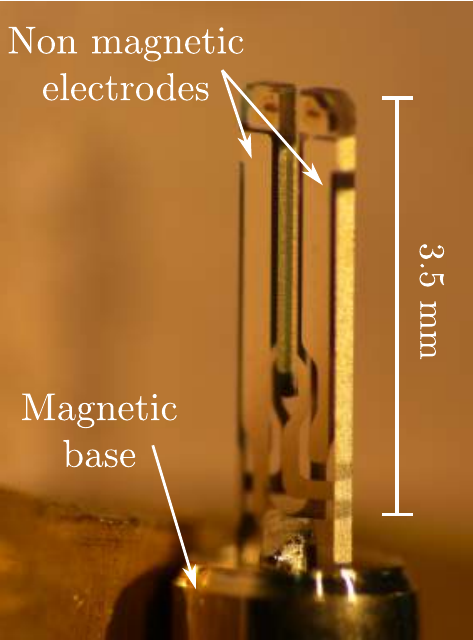}} \
 \subfloat[]{\includegraphics[height = 5cm]{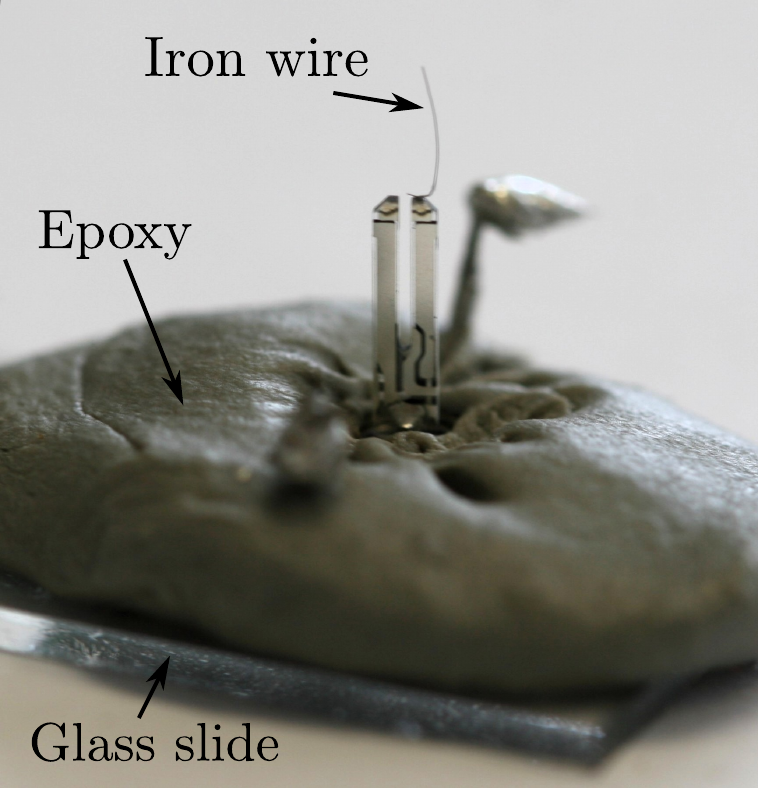}}
 \caption{(a) Picture of a TF used for performing torque differential magnetometry. The two prongs with the non-magnetic electrodes are visible at the top. The base of the quartz TF is embedded in a magnetic base visible at the bottom. (b) Picture depicting a TF cemented onto a glass slide including the magnetic base. Using this mounting technique, the unwanted magnetic field dependence is suppressed.}
\label{pictures}
\end{figure}

\begin{figure}[htb]
 \centering 
 \includegraphics[width = 8.5cm]{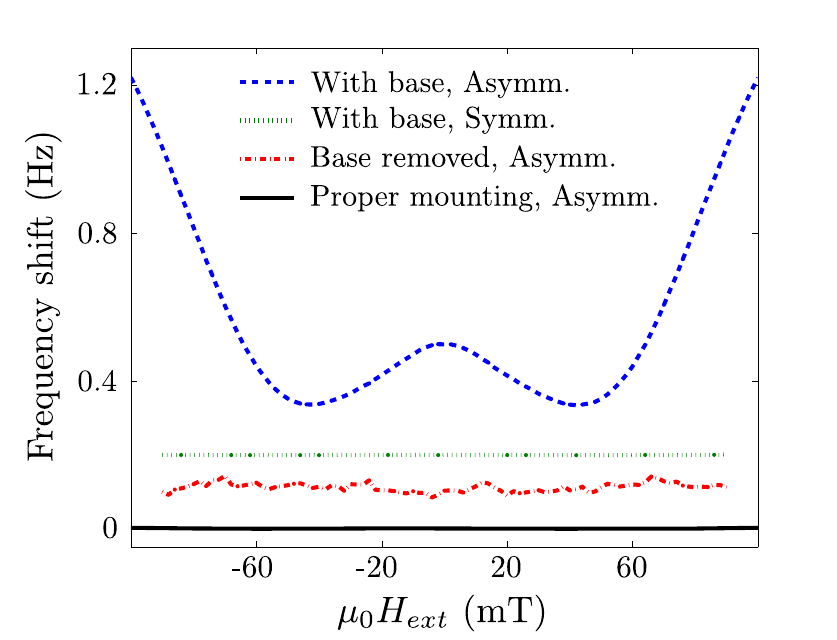}
 \caption{Resonance frequency shift vs. applied magnetic field for different mounting configurations of the TF. The label ``Asymmetric'' configuration refers to a TF mounted with a non-magnetic specimen on one prong and the ``symmetric'' configuration refers to TF operated without a specimen. For the asymmetrically loaded TF without appropriate mounting of the base, we find a strong W-shaped magnetic field dependence (blue dashed line). In contrast, the symmetrical configuration (green dotted line) shows no significant magnetic field dependence. We also show an asymmetrically loaded TF, where we have removed the base (red dash-dotted line). Here, no magnetic field dependence is observable, but the noise in the data is increased due to the lower quality factor of the TF. Mounting the TF as shown in Fig. \ref{pictures}(b) allowed for asymmetric loading while no magnetic field dependence is observed (solid black line). The curves are offset for clarity.}
 \label{confcomp}
\end{figure}

The TFs employed come in an evacuated casing which we remove to gain access to the prongs required for mounting the specimen as sketched in Fig. \ref{ctmvstdm}. To this end, we lathe off the top cap of the TF so that the prong, along with its electrical contacts, remains connected to the base (see Fig. \ref{pictures}(a)). We find, experimentally, that the casing and the base are magnetic. Note, that as long as the actual electrodes deposited on the prong are non-magnetic, which is indeed the case, this should not obstruct the experiments. To test this hypothesis, we recorded the resonance frequency of a ``symmetric'' (no specimen attached) and an ``asymmetric'' (loaded with a non-magnetic specimen) TF as a function of the applied magnetic field strength. As shown in Fig. \ref{confcomp}, the asymmetric configuration (blue dashed line) exhibits a W shaped magnetic field dependence while the symmetric configuration (green dotted 
line) shows no significant shift in the resonance frequency on changing the applied magnetic field. \footnote{In principle, the symmetric configuration does exhibit a very weak magnetic field dependence of the resonance frequency in the range of 1 mHz. But this is due to small but finite asymmetry in any TF. Hence the magnetic field dependence of resonance frequency can be used as a method to detect asymmetry in TFs.} For comparison, Fig. \ref{confcomp} also shows an asymmetrically loaded TF removed from its magnetic base showing also no magnetic field dependence.

This behavior can be understood as follows. The mirrored motion of the two prongs in the ``anti-symmetric'' resonance mode of a perfectly symmetric (both prongs identical) TF ensures that the center of mass is at rest. This implies that exciting the anti-symmetric resonance mode does not excite the center of mass motion, and vice versa. Hence this anti-symmetric mode is completely decoupled from the TF's center of mass motion~\cite{Rychenphd}. This decoupling, in a symmetric TF, prevents the resonance frequency of the anti-symmetric mode from getting affected by the forces experienced by the TF as a whole. However, the slight asymmetry induced on attaching the specimen to one prong leads to a small, but finite, coupling between the center of mass motion and the anti-symmetric resonance mode~\cite{Gomez}. Thus the resonance frequency depends, although weakly, on the net force (gradient) experienced by the TF in the presence of an applied magnetic field.

In experiments carried out at a fixed magnetic field~\cite{Rychenrsi}, this additional force provides no further complication. However, it hinders the analysis of the torque differential magnetometry data, when the TF resonance frequency needs to be recorded as function of the applied magnetic field~\cite{Todorovic,Nicks}. Thus it is desired to suppress this unwanted magnetic field dependence. To this end, we tested different configurations (including the so called qPlus configuration~\cite{Giessiblnanotech}) of TFs removed from their magnetic base and glued to a substrate. Since the packaging is part of the design for the TFs~\cite{Bottom}, some of the desirable quality criteria of the TFs are compromised on removal of the base. In particular, the robustness, reproducibility and most importantly, the high quality factor under moderate asymmetric loading are lost. The quality factor becomes sensitive to asymmetry~\cite{Gomez} and drops drastically even under smallest loads (a few micrograms). 

An alternative approach for suppressing the unwanted magnetic field dependence is to freeze the center of mass degree of freedom. We achieved this by cementing the TF including its base onto a glass slide using a two component epoxy (WIKO Alu)~\footnote{Simply gluing the TF to a substrate as in Refs. \cite{Rychenrsi,Rychenphd} was not sufficient.} [Fig. \ref{pictures} (b)]. Employing this mounting technique, the unwanted magnetic field dependence diminishes to below the measurement precision (solid black line in Fig. \ref{confcomp}). More importantly, the quality factor of the TF is practically unchanged under loadings up to a few tens of micrograms (Table \ref{qfactors}), a mass which cannot be achieved with the micrometer-sized cantilevers. Hereby, we can investigate macroscopic samples and thus extend the TF capabilities demonstrated so far~\cite{Rychenphd}.

\begin{table*}[tb]
\centering
\begin{tabular}{|c|c|c|c|c|c|c|c|}
\hline
Specimen mass & 0 $\mu$g & 17 $\mu$g & 35 $\mu$g & 52 $\mu$g & 69 $\mu$g & 87 $\mu$g & 104 $\mu$g \\ \hline
Quality factor & 8535 & 8707 & 8433 & 7237 & 7464 & 1521 & 1698 \\ \hline
\end{tabular}
\caption{Quality factors of TFs mounted as depicted in Fig. \ref{pictures}(b) for various loadings. Up to 70 $\mu$g, only a small decrease in the quality factor is observed demonstrating the possibility to investigate a wide range of magnetic specimen with this technique. The mass error in the loading is estimated at 5 $\mu$g due to the glue used to mount the specimen.}
\label{qfactors}
\end{table*}

%-------------------------------------------------------------------------------------------------------------%

\section{Measurement schemes}\label{measurement}
For the analysis of torque differential magnetometry we need to measure the resonance frequency of the mechanical resonator and its amplitude at resonance. These quantities depend on the capacitance $C$ and the resistance $R$, respectively, of the lumped $LCR$ model representing the motional branch of the BvD equivalent circuit of the TF (See Fig. \ref{ctmvstdm}). The standard method for obtaining these parameters is to measure the full frequency response at every magnetic field value and analyze the result by fitting a Lorentzian [Eq. (\ref{lorentzian})]~\cite{Stipe}. We will refer to this method as {\it Lorentzian fitting}. Nevertheless, this is not the most time efficient procedure to determine the relevant experimental parameters. To this end, we employed a phase lock loop and a pure phase detection technique as detailed in the following.  

\begin{figure}[htb]
 \centering
 \subfloat[]{\includegraphics[width = 8cm]{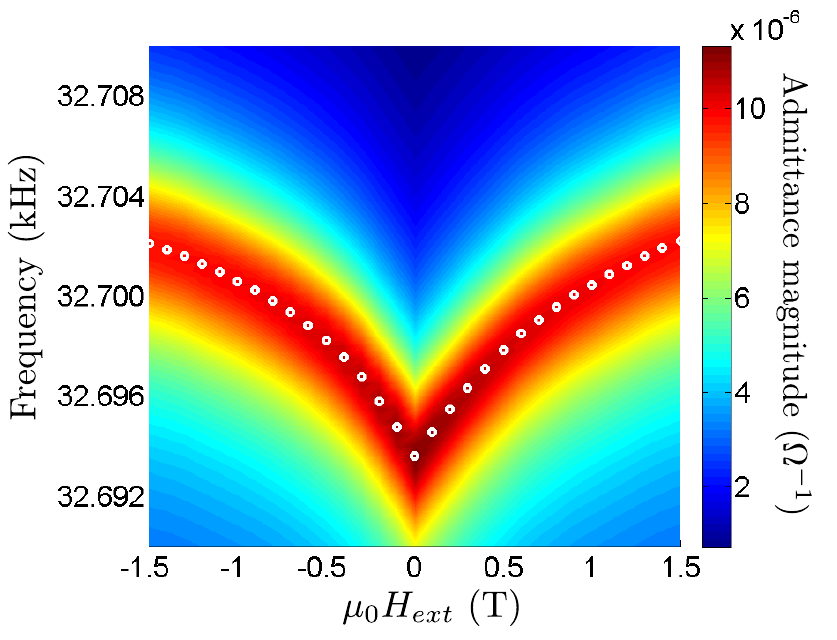}} \\
 \subfloat[]{\includegraphics[width = 8cm]{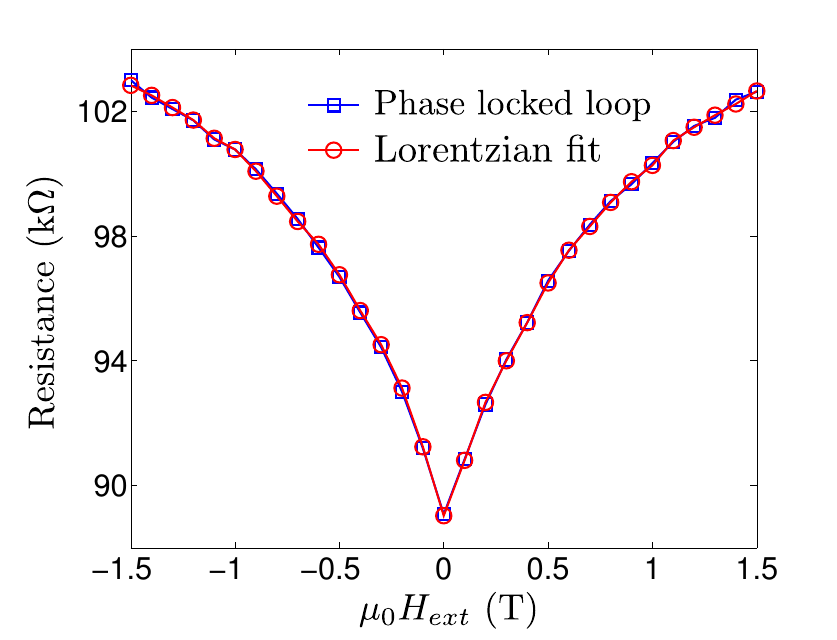}} 
 \caption{Comparison of Lorentzian fitting and PLL method for determination of resonance frequency and resistance for the iron wire specimen (see text). Both methods are found to yield the same result within the accuracy of the measurement, while the PLL method is much faster. (a) False color plot of admittance magnitude vs. frequency and applied magnetic field. The white circles denote the resonance frequency measured using the PLL method. (b) Resistance vs. applied magnetic field.}
 \label{concheck}
\end{figure}

{\it Phase-locked loop}: We have implemented a phase-locked loop~\cite{Giessiblrmp,Rychenphd} (PLL) using Labview. We begin our experiment by measuring a full admittance spectrum. Using this information, we determine the resonance frequency and the corresponding (experimental) phase. In the subsequent magnetic field sweep, the frequency of the ac drive is adjusted to maintain a constant phase that corresponds to the resonance frequency. Once the PLL has adjusted the drive frequency to the current resonance frequency, the measurement of the admittance magnitude yields information on the resistance $R$:
\begin{eqnarray}
 R & \approx & \frac{1}{|Y(f_r)|} \left(1 + \frac{\left(2 \pi f_r C_s\right)^2}{2\left| Y(f_r) \right|^2} \right). 
\end{eqnarray}
The above approximation has been obtained under the valid assumption $|Y(f_r)| \gg 2 \pi f_r C_s$, and the value of $C_s$ is determined from the initial frequency spectroscopy and Lorentzian fitting. A comparison between the PLL technique and the Lorentzian fitting method is presented in Fig. \ref{concheck} for the iron wire specimen detailed in Sec. \ref{tdmsec} (henceforth simply called the {\it iron wire specimen}). For each magnetic field value, we measured a full frequency sweep of the  admittance, which is displayed in panel a). Additionally, we determined the TF resonance frequency and resistance using the PLL method finding excellent quantitative agreement between both. 

Comparing the data acquisition speed of the experiment, we find that we require about 10 frequency points for a successful fitting of the Lorentzian lineshape. In contrast, we need only 3 to 6 admittance measurements using our PLL algorithm to obtain the same information. Thus assuming the same measurement bandwidth, we have improved the measurement speed by a factor of about two compared to the full frequency analysis. 

\begin{figure}[htb]
 \centering
 \subfloat[]{\includegraphics[width = 8.5cm]{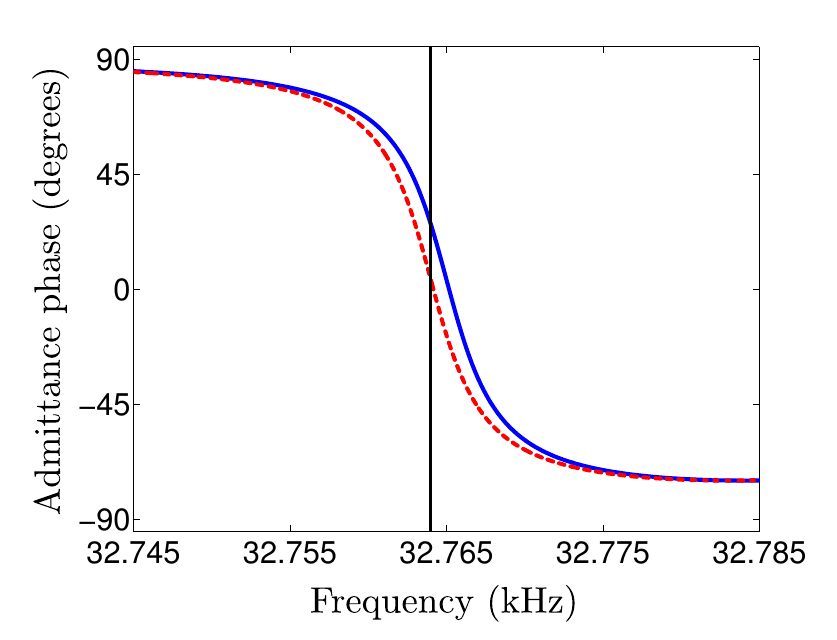}} \\
 \subfloat[]{\includegraphics[width = 8.5cm]{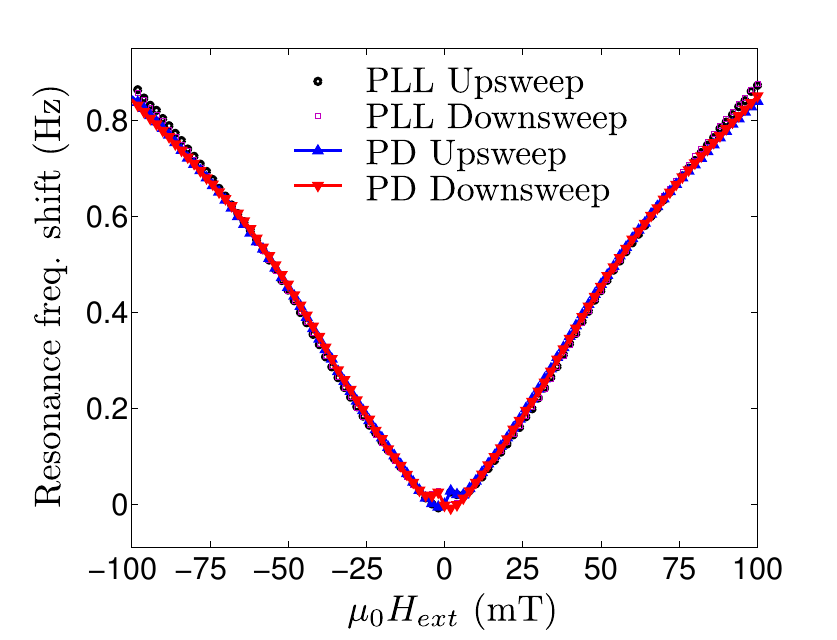}} 
 \caption{(a) Schematic of the PD method. Panel a) shows the phase response for two resonance frequencies 32764 Hz and 32765 Hz. If the linewidth is not changed significantly by the external magnetic field, Eq. (\ref{pdeq}) allows to track the resonance frequency. The black vertical line depicts the fixed drive frequency at which the phase is measured, which is then converted to the resonance frequency using Eq. (\ref{pdeq}). (b) Comparison between the phase locked loop and the phase detection technique for a torque magnetometry experiment with the iron wire specimen and the TF prepared as depicted in Fig \ref{pictures}(b). }
 \label{pd}
\end{figure}

{\it Phase detection}: By recording the phase response as a function of the applied magnetic field , it is also possible to obtain the quantities of interest within certain constraints. During a magnetic field sweep, only two parameters of the resonator ($C$ and $R$) change. If the resonator is driven at a fixed frequency (close to the resonance frequency), the admittance magnitude and phase measurements yield two equations with two unknown values. In general, these two equations need to be solved numerically. When assuming a constant resistance (hence constant $\Delta f$) and thus using only the measured phase information, we can give an analytic expression for the frequency shift. Errors are small for a high quality factor resonator since there is a steep phase change and a weak magnitude change close to the resonance frequency (see Fig. \ref{admcurves}). Using Eq. (\ref{lorentzian}) and the condition $2 \pi f_r C_s R \ll 1$, we obtain
\begin{eqnarray}\label{pdeq}
 f_r & = & f_d + \frac{\Delta f}{2} \tan (\phi),
\end{eqnarray}
where $f_d$ is the fixed drive frequency and $\phi$ the measured phase and $\Delta f$ is determined from the initial full frequency response spectroscopy. Fig. \ref{pd} shows the schematic of the phase detection (PD) scheme and a comparison between resonance frequency shift obtained by PD and PLL methods. We find a very good quantitative agreement for the two methods. Only at relatively large magnetic fields, we observe a small deviation between the two data sets stemming from our assumption of a fixed linewidth $\Delta f$ in the PD method. Note that the PD technique further reduces our measurement time by a factor of about 3-5, since a single datapoint is acquired for each magnetic field strength.

{\it Comparison between and hybrid of methods}: Clearly, Lorentzian fitting for each magnetic field value is not an efficient method. The PLL method measures the resonance frequency shift by tracking the phase at the point of highest slope, and thus highest sensitivity. The PD technique is the fastest method possible requiring only one admittance measurement but is limited to small frequency shifts.~\footnote{The choice of the best method becomes important when single admittance measurement time is limited from below by the settling time ($\sim Q/f_r$) of a very high quality factor resonator. This can easily happen at low temperature and pressure.} A hybrid of PD and PLL methods can also be employed. One can use the PD method as long as the frequency shift stays within certain acceptable range. Once the shift reaches the specified limit, one PLL step can be activated bringing the drive frequency to the current resonance frequency. After that the PD method can take over once again.

%--------------------------------------------------------------------------------------------------------------%

\section{Torque differential magnetometry}\label{tdmsec}

In this section, we discuss the proof of principle experimental results of torque differential magnetometry performed on a sample with uniaxial shape anisotropy. The specimen is chosen for a direct comparison of the data and the experimental method with Refs. \cite{Stipe} and \cite{Weber}. Our setup consists of a $3.5 \pm 0.5$ mm long and $25 \pm 2.5~\mu$m diameter iron wire (mass $\sim 13 ~\mu$g) attached to a TF as shown in Fig \ref{pictures}(b). Such a thin magnetic wire is known to have a uniaxial shape anisotropy~\cite{Chikazumi} and can be described within the Stoner-Wohlfarth single domain approximation~\cite{Chikazumi} via the free energy density $F = K_u \sin^2(\theta)$, where $\theta$ is the angle between the magnetization and anisotropy axis, and $K_u (> 0)$ is the anisotropy constant. 

\begin{figure}[htb]
 \centering
 \includegraphics[height = 6cm]{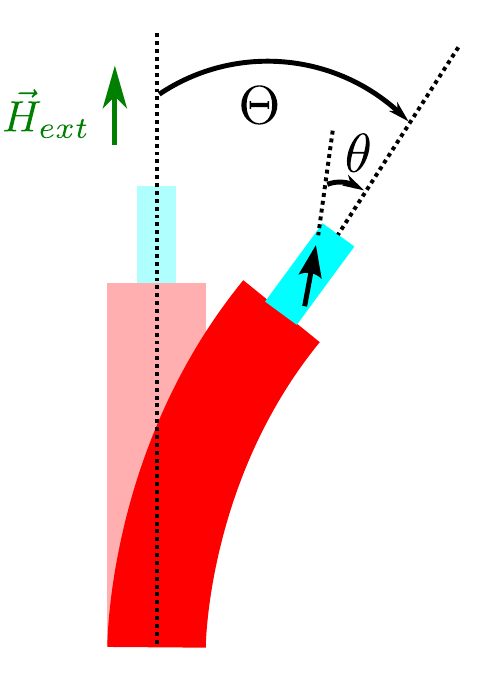}
 \caption{Schematic depicting a single prong of the TF (red) loaded with a magnetic specimen (light blue) displaced from equilibrium. The uniaxial anisotropy direction deviates by an angle $\theta$ from the magnetization direction and an angle $\Theta$ from the applied magnetic field. }
 \label{canting}
\end{figure}

\begin{figure}[htb]
 \centering
 \subfloat[]{\includegraphics[width = 8.5cm]{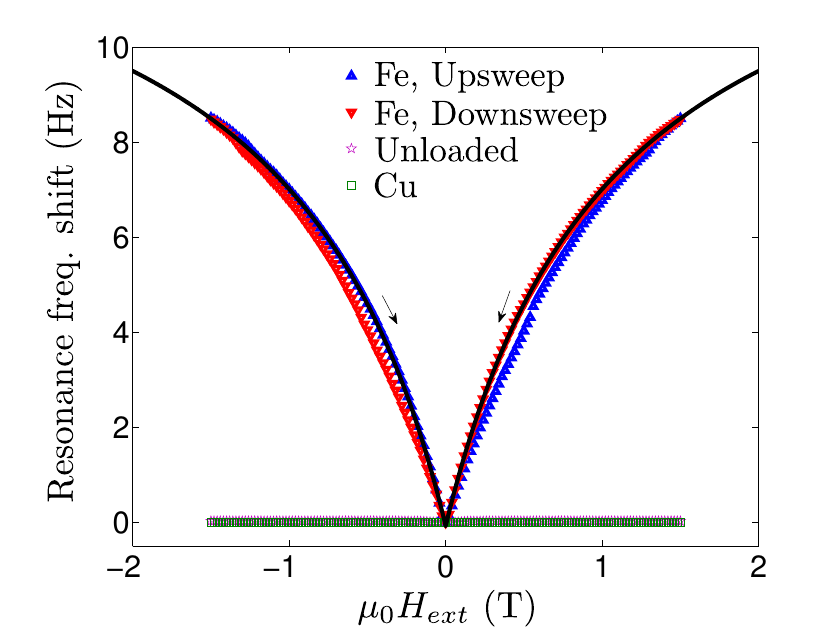}} \\
 \subfloat[]{\includegraphics[width = 8.5cm]{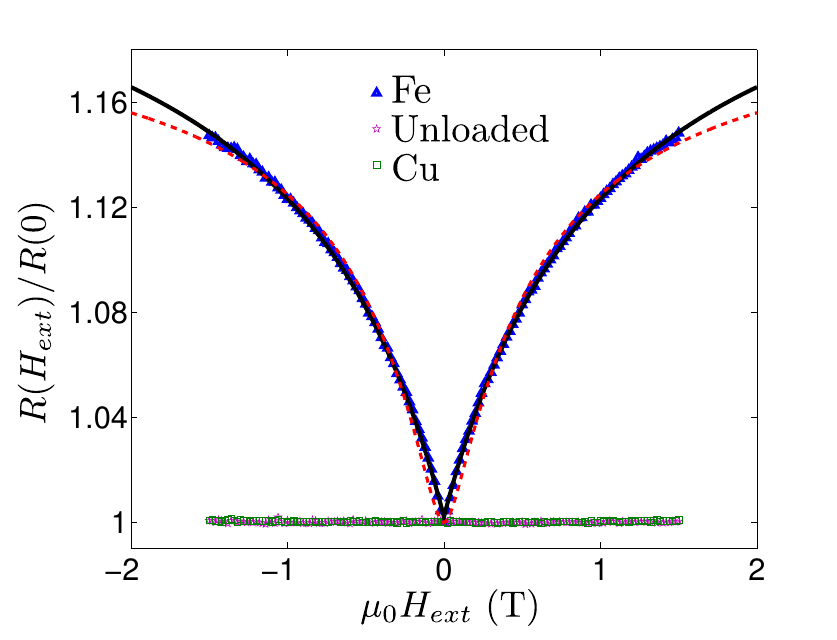}} 
 \caption{Magnetic field dependent resonance frequency shift (a) and resistance (b) for the iron wire specimen [Fig. \ref{pictures}(b)] (triangles), no specimen (stars), and a non-magnetic Copper specimen (square symbols). All data are unaveraged and show a single magnetic field sweep. (a) The fit of the resonance frequency shift (solid black line) to Eq. (\ref{freqshift}) yields an anisotropy field of $1.08 \pm 0.01$ T in good agreement with the value expected from the shape anisotropy. (b) The fit of resistance data to $c_1 [ H_{\textrm{ext}}/ ( H_{\textrm{ext}} + H_k) ]$ (solid black line) yields an almost perfect fit and $\mu_0 H_k = 1.07 \pm 0.03$ T in consistence with the frequency shift data, while a fit to $c_2 [ H_{\textrm{ext}}/ ( H_{\textrm{ext}} + H_k) ]^2$ (red dashed line) yields a bad fit and inconsistent value of $\mu_0 H_k = 0.28 \pm 0.03$ T.}
 \label{tdmfig}
\end{figure}

We consider an applied magnetic field collinear with the anisotropy easy axis under equilibrium orientation of the prong. If the prong is deflected by a small angle $\Theta$ from this direction, the magnetic moment of the specimen makes an angle $\theta = [ H_{\textrm{ext}}/ ( H_{\textrm{ext}} + H_k) ] \Theta$ with the anisotropy axis (Fig. \ref{canting}), where $\mu_0 H_k = 2 K_u/M_s$ is the anisotropy field, with the saturation magnetization $M_s$~\cite{Stipe,Kamra}. Alternately, the angle between the magnetic moment and the applied magnetic field direction is $\Theta - \theta$. This gives rise to an additional restoring torque~\cite{Stipe,Kamra} $\tau_m = M_s V \mu_0 H_{\textrm{ext}} (\Theta - \theta) = M_s V \mu_0 H_{\textrm{ext}} [H_k/H_{\textrm{ext}} + H_k] \Theta$, and an additional effective stiffness of $k_m  \ = \ \tau_m/\Theta L_e^2 \ = \ (M_s V/L_e^2 )\mu_0 H_{\textrm{ext}} [H_k/H_{\textrm{ext}} + H_k] $, where $V$ is the volume of the specimen, and $L_e$ is the effective length of the prong~\cite{Morse}. 
Under the condition $k_m \ll k_{\textrm{el}}$ corresponding to a weak perturbation 
of the resonance frequency, the resonance 
frequency shift becomes:
\begin{eqnarray}\label{freqshift}
 \delta f_r(H_{\textrm{ext}}) & = & \frac{f_{\textrm{el}}M_s V \mu_0}{2 k_{\textrm{el}} L_e^2} \ \frac{|H_{\textrm{ext}}| H_k}{|H_{\textrm{ext}}| + H_k},
\end{eqnarray}
where $f_{\textrm{el}}$ and $k_{\textrm{el}}$ are respectively the resonance frequency and effective stiffness of the oscillator at zero applied magnetic field. The absolute value of $H_{\textrm{ext}}$ appears because the magnetization too reverses its direction on reversal of magnetic field direction. A fit of the frequency shift data to Eq. (\ref{freqshift}) yields the anisotropy field of the specimen as $\mu_0 H_k = 1.08 \pm 0.01$ T [see Fig. \ref{tdmfig}(a)]. A clear hysteresis in the resonance frequency shift vs. magnetic field curve is seen. The low field curve is depicted in Fig. \ref{pd}(b) where hysteretic kinks can be seen close to zero field region. Assuming that the anisotropy is purely due to the shape of the specimen, the saturation magnetization becomes $\mu_0 M_s = 2.16 \pm 0.02$ T in good agreement with the literature value for iron 2.15 T~\cite{Chikazumi}.

The magnetic field dependent resistance is shown in Fig. \ref{tdmfig} (b). The resistance curve does not show a strong hysteresis consistent with the observation of Stipe and co-workers~\cite{Stipe}. However, we find the magnetic contribution to the resistance proportional to the magnetization canting amplitude ($\theta_{max} = [ |H_{\textrm{ext}}|/ ( |H_{\textrm{ext}}| + H_k) ] \Theta_{max}$). This is in contrast with the observation of Stipe {\it et al.}~\cite{Stipe}, who find the magnetic contribution to the dissipation coefficient to be proportional to the square of $\theta_{max}$. We attribute this observation to the slight deviation of our wire from the optimal alignment with respect to the applied magnetic field. Here, we expect that the linear effects dominate. For observing the quadratic effects, a perfect alignment of the sample with respect to the applied magnetic field is crucial. We notice that our data on the magnetic field dependent dissipation coefficient has little spread as compared to the 
data reported by 
Stipe and co-workers~\cite{Stipe}.  Weber {\it et al.}~\cite{Weber}, on the other hand, were not able to resolve the dependence of dissipation coefficient on the applied magnetic field in their measurements. Hence our technique looks particularly promising for the investigation of magnetic dissipation. 

The data depicted in Fig. \ref{tdmfig}, which is a single full magnetic sweep recorded without any averaging, underlines the high signal-to-noise ratio in our measurement. The plotted data for a single specimen was captured within 15 minutes, although several magnetic field sweeps were recorded to check their reproducibility. A comparison between Fig. \ref{tdmfig} and the corresponding plots in Ref. \cite{Stipe}, in particular the data for dissipation coefficients, makes the advantage of our measurement scheme evident.  

%---------------------------------------------------------------------------------------------------------%

\section{Discussion}\label{discussion}

We have demonstrated an inexpensive and all electrical setup for torque differential magnetometry~\cite{Stipe,Weber,Kamra} using piezoelectric quartz tuning forks (TFs) capable of operation over a broad temperature and pressure range. A high signal-to-noise ratio under ambient conditions was achieved at a lock-in effective bandwidth of about 1 Hz. The anisotropy field and saturation magnetization of an iron wire specimen were quantitatively extracted corroborating the literature values. We also demonstrated the possibility to measure specimens with a mass of up to $\sim 70~ \mu$g without any significant loss in sensitivity, and a high sensitivity for detecting magnetic contribution to dissipation.

In the following we estimate the minimum mass of the same iron wire that could be used in our measurements (leaving all parameters unchanged) with a signal-to-noise ratio of about 1. We find that the slope of the phase vs frequency curve for our TFs (under ambient conditions) at resonance is about 35 degrees per Hz. At the bandwidth of our measurements, the phase measurement noise was found to be below 1 degree. This allows us a frequency sensitivity of below 30 mHz. In Fig. \ref{tdmfig}(a), the typical frequency shift between adjacent magnetic field values (separated by 20 mT) is about 200 mHz. This implies that we can detect a signal that is weaker by a factor of about $200/30 \sim 7$, which corresponds to a sample weighing 7 times less than the sample investigated herein ($\sim 13~ \mu$g). Hence a specimen with mass $\sim 2~\mu$g can easily be measured using the same set of parameters that we employed for our measurement reported herein (15 minutes of measurement time). 

We notice that the extraction of the anisotropy constant requires fitting the resonance frequency shift data which need not be separated by 20 mT. Hence the more meaningful limit for the minimum specimen mass in an anisotropy parameter extraction experiment is obtained by noticing that the maximum frequency shift in Fig. \ref{tdmfig}(a) is of the order of 10 Hz. Since we can detect frequencies shifts as low as 30 mHz as argued above, it should be possible to characterize a specimen $10\,\textrm{Hz}/30\,\textrm{mHz} \sim 100$ times smaller than the one characterized here setting the limit to about $100$ ng of iron.  

Torque differential magnetometry (TDM) is a very powerful tool for the characterization of magnetic specimens, and the investigation of magnetic switching, magnetic phase transitions and high Tc superconductors (see Ref. \cite{Kamra} and references therein). As compared to superconducting quantum interference device magnetometry, TDM has a faster response time~\cite{Brugger} enabling investigation of dynamic phenomena while offering a comparable sensitivity~\cite{Stipe,Brugger}. The simple and inexpensive TDM setup demonstrated herein makes it still more attractive as a magnetometry technique.

In the present work, we employed piezoelectric quartz TFs and achieved a high sensitivity under ambient conditions. There is still a large room for gain in sensitivity by using smaller TFs at low temperature and pressure~\cite{Giessiblnanotech}. Although the present work lays emphasis on TDM, the generic measurement scheme that has been demonstrated here can be adapted for other kinds of measurements which have traditionally been done using cantilevers and optical interferometers in vacuum. In particular, our setup design eliminates unwanted artifacts due to magnetic TF base that may interfere with magnetic field dependent measurements~\cite{Rychenrsi,Rychenphd}.

%-------------------------------------------------------------------------------------------------------------%

\begin{acknowledgement}
We thank Sibylle Meyer for help with measurements in the cryostat, and Matthias Pernpeintner for fruitful discussions. Financial support from the DFG via SPP 1538 ``Spin Caloric Transport'', Project No. GO 944/4 and the Dutch FOM Foundation is gratefully acknowledged.
\end{acknowledgement}

%------------------------------------------------------------------------------------------------------------------%

\bibliographystyle{epj}
\bibliography{TDMTF}

\end{document}